\documentclass[fleqn,twoside]{article}
\usepackage{espcrc2}

\usepackage{graphicx}
\usepackage{epsfig}
\let\cl=\centerline
\def\figbox#1;#2;{\parbox{#2cm}{%
\vglue3mm\epsfig{file=#1.eps,width=#2cm}\vglue3mm}}
\def\figboxc#1;#2;{\cl{\figbox #1;#2;}}
\usepackage[figuresright]{rotating}

\newcommand{\AmS}{{\protect\the\textfont2
  A\kern-.1667em\lower.5ex\hbox{M}\kern-.125emS}}

\title{\bf{Charged kaon lifetime at KLOE}}

\author{KLOE collaboration:
F.~Ambrosino,
A.~Antonelli, 
M.~Antonelli, 
F.~Archilli,
C.~Bacci,
P.~Beltrame,
G.~Bencivenni, 
S.~Bertolucci, 
C.~Bini, 
C.~Bloise, 
S.~Bocchetta, 
V.~Bocci,
F.~Bossi,
P.~Branchini,
R.~Caloi,
P.~Campana, 
G.~Capon, 
T.~Capussela,
F.~Ceradini,
S.~Chi,
G.~Chiefari, 
P.~Ciambrone,
E.~De~Lucia,
A.~De~Santis, 
P.~De~Simone, 
G.~De~Zorzi,
A.~Denig,
A.~Di~Domenico,
C.~Di~Donato,
S.~Di~Falco,
B.~Di~Micco,
A.~Doria,
M.~Dreucci,
G.~Felici, 
A.~Ferrari,
M.~L.~Ferrer, 
G.~Finocchiaro,
S.~Fiore,
C.~Forti,       
P.~Franzini,
C.~Gatti,      
P.~Gauzzi,
S.~Giovannella,
E.~Gorini, 
E.~Graziani,
M.~Incagli,
W.~Kluge,
V.~Kulikov,
F.~Lacava, 
G.~Lanfranchi, 
J.~Lee-Franzini,
D.~Leone,
M.~Martini,
P.~Massarotti$^{\dagger}$,
W.~Mei,
S.~Meola,
S.~Miscetti, 
M.~Moulson,
S.~M\"uller,
F.~Murtas, 
M.~Napolitano,
F.~Nguyen,
M.~Palutan,          
E.~Pasqualucci,
A.~Passeri,  
V.~Patera,
F.~Perfetto,
M.~Primavera,
P.~Santangelo,
G.~Saracino,
B.~Sciascia,
A.~Sciubba,
F.~Scuri, 
I.~Sfiligoi,     
T.~Spadaro,
M.~Testa,
L.~Tortora, 
P.~Valente,
B.~Valeriani,
G.~Venanzoni,
R.~Versaci,
G.~Xu.
}
\begin{document}
\begin{center}
\begin{abstract}
\noindent
    {\bf Abstract}\\
Preliminary result on the charged kaon lifetime $\tau^{\pm}$, obtained by the 
KLOE experiment operating at DA$\Phi$NE, the Frascati $\phi$-factory, is 
presented.
\vspace{1pc}
\end{abstract}
\end{center}
\maketitle

%%%%%%%%%%%%%%%%%%%%%%%%%%%%%%%%%%%%%%%%%%%%%%%%%%%%%%%%%%%%%%%
\section{DA$\Phi$NE and KLOE}
\noindent 
The DA$\Phi$NE e$^+$e$^-$ collider operates at a total energy 
W = 1020 MeV, the mass of the $\phi$(1020)-meson.
Approximately $3\times10^6$ $\phi$ -mesons are produced for 
each integradet luminosity of 1 pb$^{-1}$.
Since 2001, KLOE has collected an integrated luminosity of 
about 2.5 fb$^{-1}$.
Results presented below are based on 2001-02 data for about 450~pb$^{-1}$.
The KLOE detector consists of a large cylindrical drift chamber, DC, surrounded
by a lead/scintillating-fiber electromagnetic calorimeter, EMC. 
The drift chamber \cite{bib:dc}, is 4~m in diameter and 3.3~m long.
The momentum resolution is $\sigma(p_{T})/p_{T} \sim 0.4\%$. 
Two track vertices are reconstructed with a spatial resolution 
of $\sim$ 3 mm. 
The calorimeter \cite{bib:emc}, composed of a barrel and two endcaps,
covers 98\% of the solid angle. 
Energy and time resolution are $\sigma(E)/E = 5.7\%/\sqrt{E[{\rm GeV}]}$ and
$\sigma(t) = 57 {\rm ps}/ \sqrt{E[{\rm GeV}]} \oplus 100 {\rm ps}$.
A superconducting coil around the detector provides a 0.52~T magnetic
field.
The KLOE trigger \cite{bib:trg}, uses calorimeter and drift chamber
information. 
For the present analysis only the electromagnetic calorimeter (EMC) 
signals have been used. Two local energy deposits above threshold, 
$E_{\rm th}>50$ MeV for the barrel and $E_{\rm th}>150$ MeV for the 
endcaps, are required.

%%%%%%%%%%%%%%%%%%%%%%%%%%%%%%%%%%%%%%%%%%%%%%%%%%%%%%%%%%%%%%%
\section{The tag mechanism}
\begin{figure}[htb]
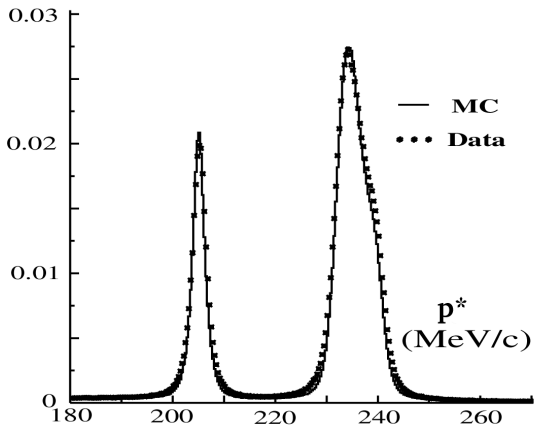

\figboxc tag;7;\vglue-.8cm
%\vspace{9pt}
\caption{\footnotesize{Momentum spectrum in the kaon rest frame of the negative charged
         decay particle, assuming the particle has the pion mass for data 
	 (dots) and MC (lines).The distribution are normalized to
         unity. The two peaks correspond to pions and muons 
	 from $K^- \rightarrow \pi^- \pi^0$ (205 MeV/c) and $K^-
         \rightarrow \mu^- \nu_\mu$ (236 MeV/c). The muon peak is
         broadened by the use of the incorrect mass}}
\label{fig:tag_spectrum}
\end{figure}

Most of the case in its center of mass $\phi$-mesons decay into 
anti-collinear $K\bar{K}$ pairs. 
In the laboratory this remains approximately true because of the small
crossing angle of the e$^+$ and e$^-$ beams.
Therefore the detection of a $K(\bar K)$ tags the presence of a  
$\bar K (K)$ of given momentum and direction. 
%This is called tag mechanism.
The decay products of the $K^{\pm}$ pair define two spatially well
separated regions called the tag and the signal hemispeheres.
Identified $K^{\mp}$ decays tag a $K^{\pm}$ beam and provide sample 
count, using the total number of tags as normalization. 
This procedure is a unique feature of a $\phi$-factory and provides the
means for measuring absolute branching ratios. 
%i.e. ratios $\Gamma_i / \Gamma_{tot}$ rather than 
%ratios of BR's $\Gamma_i / \Gamma_{j}$.
Charged kaons are tagged using the two body decays 
$K^{\pm}\rightarrow \mu^\pm\rlap{\raise1.2ex\hbox{\scriptsize($-$)}}
\kern.3em\nu_{\,\mu}$ and $K^{\pm}\rightarrow \pi^{\pm} \pi^0$. 
Since the two body decays correspond to about 85\% of the charged kaon
decays \cite{bib:pdg} and since $BR(\phi \rightarrow K^+K^-)\simeq 49\%$ 
\cite{bib:pdg}, there are about  $1.5 \times 10^6 K^+K^-$ events/pb$^{-1}$. 
The two body decays are identified as peaks in the momentum spectrum of the
secondary tracks in the kaon rest frame and computed assuming $m_\pi$ for
the particle (Fig. \ref{fig:tag_spectrum}).
In order to minimize the impact of the trigger efficiency, the taging kaon 
must provide the EMC trigger of the event, so called self-triggering
tags. 
$N_{\rm selftrg\  tag} \approx 2 \times 10^5$ per pb$^{-1}$. 
\vglue-5mm

%%%%%%%%%%%%%%%%%%%%%%%%%%%%%%%%%%%%%%%%%%%%%%%%%%%%%%%%%%%%%%%
\section{Measurement of the charged kaon lifetime}
\noindent
The measurement is performed using 230 pb$^{-1}$ collected at $\phi$ peak.
The data sample has been split in two uncorrelated subsamples, 
150 pb$^{-1}$ have been used for the measurement, the remaining 
80 pb$^{-1}$ have been used to evaluate the efficiencies.
$K_{\mu 2}$ tags of both charges have been used.
There are two methods available for the measurement: the kaon decay length
and the kaon decay time.
The two methods allow cross checks and studies of systematics; their
resolutions are comparable.
The method relying on the measurement of the charged kaon decay length
requires first the reconstruction of the kaon decay vertex in the
fiducial volume  using only DC information: the signal is given by a 
$K^\pm$, moving outwards in the DC with momentum
$70 < p_K < 130$ MeV/c and having point of closest approach to the
interaction point (IP) with $0 < \sqrt{x^2_{PCA} +y^2_{PCA}} < 10$ cm 
and $|z_{PCA}| < 20$ cm. The kaon decay vertex in the DC fiducial volume
($40 < \sqrt{x^2_V +y^2_V} < 150$ cm, $|z_V| < 150$ cm) is required.
Once the decay vertex has been identified the kaon track is extrapolated
backward to the interaction point into 2 mm steps, taking into account the 
ionization energy loss $dE/dx$ to evaluate its velocity $\beta c$.
Then the proper time is obtained from:
\begin{equation}
t^* = \sum_i \Delta t_i = 
\sum_i \frac{\sqrt{1-\beta^2_i}}{\beta_i} \Delta l_i
\end{equation}
The efficiency has been evaluated directly from data.
The control sample has been selected using calorimetric information only, 
selecting for a neutral vertex: two clusters in time fired by the photons
coming from the $\pi^0$ decay.
The proper time is fitted between 16 and 30 ns correcting for the
efficiency.
Resolution effects have been taken into account.
The preliminary result we have obtained, which is the weighted mean between 
 the $K^+$ and the $K^-$ lifetimes, is:
\begin{center}
\begin{equation}
\tau^\pm = (12.367 \pm 0.044 \pm 0.065)\ ns
\end{equation}
\end{center}
The evaluation of systematic uncertainties is sill preliminary, 
final numbers will be presented at the conference.
\begin{figure}[htb]
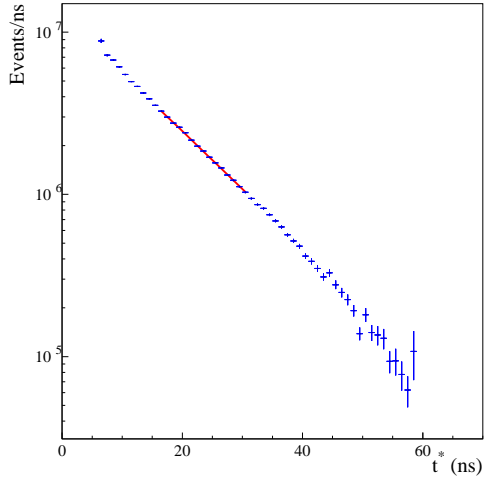

\vglue-8mm\figboxc fit_length_new;6.5;\vglue-10mm
\caption{\footnotesize{Charged kaon proper time distribution, obtained with the first method, fitted (red line) with a convolution of an exponential and a resolution function}}
\label{fig:tau_spectrum}
\end{figure}\vglue-5mm
The second method relies on the measurement of the kaon decay time. We consider only events with a $\pi^0$ in the final state:
\begin{equation}
K^\pm \rightarrow X + \pi^0 \rightarrow  X + \gamma \gamma
\end{equation}
We can obtain the kaon time of flight using the time ot the EMC clusters of the photons from the $\pi^0$ decay. We require the backward extrapolation to the interaction point of the tagging kaon track and the forward extrapolation of the helix of the other kaon on the signal side. Stepping along the helix we look for the $\pi^0\rightarrow \gamma 
\gamma$ decay vertex without looking at the real kaon track. For each photon it is possible to measure the kaon proper decay time
\begin{equation}
t^* = (t_\gamma - \frac{r_\gamma}{c} - t_\phi) \cdot \sqrt{1-\beta^2_K}
\end{equation}
The efficiency has been evaluated directly from data.
The control sample has been selected using drift chamber information only, 
selecting the kaon decay vertex in the fiducial volume.
The proper time is fitted between 13 and 42 ns correcting for the
efficiency. Resolution effects have been taken into account.
The weighted mean between the $K^+$ and the $K^-$ lifetimes gives as 
preliminary result:
\begin{equation}
\tau^\pm = (12.391 \pm 0.049 \pm 0.025)\ ns
\end{equation}
\begin{figure}[htb]
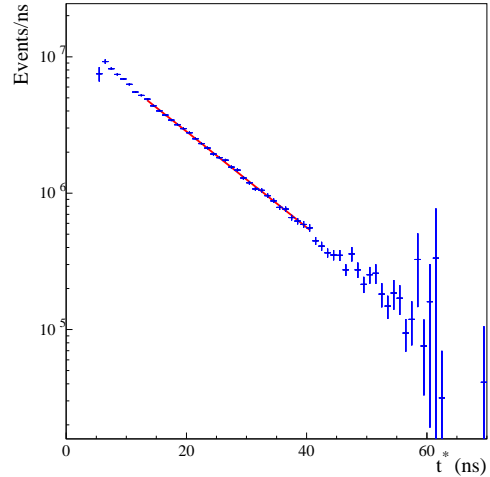

\vglue-8mm\figboxc fit_time_new;6.5;\vglue-10mm
\caption{\footnotesize{Charged kaon proper time distribution, obtained with the second method, fitted (red line) with a convolution of an exponential and a resolution function}}
\label{fig:time_spectrum}
\end{figure}
The evaluation of systematic uncertainties is sill preliminary, final numbers will be presented at the conference. In order to evaluate the statistical correlation between the two methods we divide the data sample into five subsamples. For each subsample, and for each method, we evaluate the proper time distribution and its efficiency. The value of the correlation is
\begin{equation}
\rho = .338
\end{equation}
The weighted mean between the two charges and between the two methods is 
\begin{center}
\begin{equation}
\tau^\pm = (12.384 \pm 0.048)\ ns
\end{equation}
\end{center}
%%%%%%%%%%%%%%%%%%%%%%%%%%%%%%%%%%%%%%%%%%%%%%%%%%%%%%%%%%%%%%%

\end{document}